\documentclass[a4paper,superscriptaddress,prl,twocolumn]{revtex4}
\usepackage{graphicx}

\renewcommand{\Re}{\mbox{Re}}
\newcommand{\Ra}{\mbox{Ra}}
\newcommand{\Rac}{\mbox{Ra}_{c}}
\renewcommand{\Pr}{\mbox{Pr}}
\newcommand{\Nu}{\mbox{Nu}}

\begin{document}

\title{Wind reversals in turbulent Rayleigh-B\'enard convection}

\author{Francisco \surname{Fontenele Araujo}}
\affiliation{Department of Applied Physics and J. M. Burgers Centre
for Fluid Dynamics, University of Twente, 7500 AE Enschede, The Netherlands}
\author{Siegfried Grossmann}
\affiliation{Department of Physics, University of Marburg, Renthof 6, D-3502 Marburg, Germany}
\author{Detlef Lohse}
\affiliation{Department of Applied Physics and J. M. Burgers Centre
for Fluid Dynamics, University of Twente, 7500 AE Enschede, The Netherlands}

\begin{abstract}
The phenomenon of irregular cessation and subsequent reversal
of the large-scale circulation in turbulent Rayleigh-B\'enard
convection is theoretically analysed. The force and thermal
balance on a single plume detached from the thermal
boundary layer yields a set of coupled nonlinear equations,
whose dynamics is related to the Lorenz equations.
For Prandtl and Rayleigh numbers in the range
$10^{-2} \leq \Pr \leq 10^{3}$
and
$10^{7} \leq \Ra \leq 10^{12}$,
the model has the following features:
(i) chaotic reversals may be exhibited at Ra $\geq 10^{7}$;
(ii) the Reynolds number based on the root mean square velocity
scales as
$\Re_{rms} \sim \Ra^{[0.41\, \cdots \,0.47]}$
(depending on Pr),
and as
$\Re_{rms} \sim \Pr^{-[0.66\,\cdots\,0.76]}$
(depending on Ra);
and
(iii) the mean reversal frequency follows an effective scaling law
$\omega / (\nu \, L^{-2}) \sim \Pr^{-(0.64 \pm 0.01)}\,\Ra^{0.44 \pm 0.01}$.
The phase diagram of the model is sketched, and the observed transitions
are discussed.
\end{abstract}

\date{\today}


\maketitle

One important issue in turbulent Rayleigh-B\'enard convection
is the interplay between the large-scale circulation (the so-called wind)
\cite{Krishnamurti/wind/PNAS/1981} and the dynamics of plumes detached
from the thermal boundary layers \cite{Xia/wind/onset/JFM/2004}.
In particular, such interplay seems to be relevant in the process of
circulation reversals, which occur in an irregular time sequence
\cite{Benzi/PRL/2005,Sreenivasan/wind/PRE/2002,Niemela/wind/JFM/2001,Cioni/JFM/1997,sano05,ahlers}.
Remarkably, similar reversals are also observed in the wind
direction of the atmosphere \cite{Sreenivasan/wind/PoF/2000}
and in the magnetic polarity of the earth
\cite{Glatzmeier/geomagnetic/Nature/1999}.

In principle, two reversal scenarios are possible:
Reversal through cessation of the convection roll,
and reversal through its azimuthal rotation. With
two temperature sensors placed close to each other
near the sidewall \cite{Niemela/wind/JFM/2001,Sreenivasan/wind/PRE/2002},
one can detect roll reversals, but not distinguish between
the two scenarios. With several sensors placed along
the azimuth of the cell,  Cioni {\it et al.} \cite{Cioni/JFM/1997}
succeeded to detect reversal through azimuthal rotation of the roll.
Reversal through rotation was also detected in refs.\ \cite{sano05,ahlers}.
However, with an ingenious multi-probe setup, Brown, Nikolaenko, and Ahlers
\cite{ahlers} were able to distinguish between the rotation and cessation scenarios,
and many reversals through cessation were detected.
Reversal through cessation was also observed in {\it two-dimensional}
numerical simulations of the Boussinesq equations
(see fig.\ 8 of ref.\ \cite{Hansen/1992} and fig.\ 12 of ref.\
\cite{Furukawa/Onuki/PRE/2002}),
where the rotation scenario is of course impossible.

Since reversal through cessation is a more surprising scenario,
the aim of the present work is to reveal its physical mechanism.
Qualitatively, the picture is as follows \cite{Grossmann/Ilmenau}:
If an uprising hot plume gets too fast because of a temperature
surplus, it fails to cool down sufficiently when passing the top
plate. It then is still warmer than the ambient fluid when advected
down along the sidewall. By buoyancy it therefore looses speed and
counteracts the large-scale circulation. Indeed, the downward wind
may be counteracted so strongly that it stops or even reverses its
direction. This mechanism can be effective only for sufficiently
strong wind, i.e., for sufficiently large Reynolds number,
because for slow motion the thermal diffusivity $\kappa$ has
enough time to reduce the temperature surplus of the originally
warmer plume relative to its neighbourhood. Then its power
to reverse the circulation by buoyancy is gone.

\begin{figure}[htbp]
\begin{center}
\includegraphics*[height=4cm]{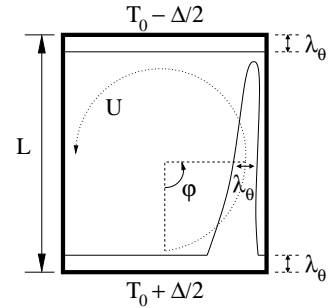}
\caption{\label{rb-cell}
Sketch of the motion of a single plume of width $\lambda_{\theta}$
and height $L$. In an aspect-ratio-one container, the circulation
radius is given by $L/2$.}
%\vspace{-0.7cm}
\end{center}
\end{figure}

\textit{The model}: In order to quantify the cessation mechanism discussed
above, let us first characterize the size of a circulating plume. As shown in figure
\ref{rb-cell}, a single plume will be understood as a thermal structure
of width $\lambda_{\theta}$ (the thickness of the thermal boudary-layer
from which it originated) and length $L$ (the height of the convection
container). In addition, its volume is assumed to scale as
$\lambda_{\theta}^{2}\,L$, with a typical cross-section area $\lambda_{\theta}^{2}$,
and surface area $\lambda_{\theta}\,L$.

Supposing that such a plume circulates with velocity $U(t)$,
it is reasonable to expect that its dynamics is essentially a matter
of balance between buoyancy and drag.

In the Boussinesq approximation the buoyancy force
(per mass) is given by
$\mbox{\textbf{f}}_{b}
=
- \,\alpha_{p}\,(T - T_{0})\,\mbox{\textbf{g}},
$
where
$\alpha_{p}$ is the isobaric thermal expansion coefficient,
$T$ the plume temperature,
$T_{0}$ the mean  temperature, and
$g$ the gravitational acceleration.
On the other hand, the drag force (per mass) on the
plume has the strength
$
f_{d}
=
\frac{1}{2}\,C(\Re) \,U^{2}\,L^{-1},
$
where $C(\Re)$ is the drag coefficient, and
the Reynolds number is defined by $\Re=L\,U\,\nu^{-1}$.
Here, $C(\Re)$ is taken as \cite{Lohse/crossover}:
\begin{equation}
\label{drag}
C(\Re)
=
\sqrt{\left(\frac{6}{b}\right)^{3}}
\left[
\sqrt{\frac{3b^{3}}{8}}\,\frac{1}{\Re}
+
\sqrt{1+\frac{3b^{3}}{8\,\Re^{2}}} \,\,
\right],
\end{equation}
where $b = 8.4$ is the Kolmogorov constant. Equation (\ref{drag})
describes the transition from the strongly decreasing drag $\sim 1/\Re$
in the viscous regime to the Re-independent drag in the turbulent regime.
As pointed out in ref. \cite{Lohse/crossover}, it pretty well agrees
with experimental data.

Now, let us consider the thermal interaction
between a single plume and its surrounding.
Strictly speaking, the surrounding consists of the fluid as
well as the sidewalls, and the top and bottom plates. We do
not distinguish between all these and describe the temperature
of the plume surrounding $T_{s}(\varphi)$ by a time-independent profile:
\begin{eqnarray}
\label{T_surrounding}
T_{s}(\varphi)&=& T_{0} + \frac{\Delta}{2}\cos \varphi,
\end{eqnarray}
where $\Delta$ is the temperature difference between the
horizontal plates. We Fourier expand the temperature
variable of the plume:
\begin{equation}
\label{T}
T(\varphi,t) =
T_{0}
+
\sum_{n=1}^{\infty}
\left[
A_{n}(t)\, \sin(n\varphi) + B_{n}(t)\cos(n\varphi)
\right],
\end{equation}
where $A_{n}(t)$ and $B_{n}(t)$ are the amplitudes.

\textit{Equations of motion}:
In order to derive the equations of motion for a single plume, we
follow an analogy with the Malkus waterwheel \cite{Strogatz,Kolar/PRA/1992}.
%a mechanical system whose rotation may also exhibit reversals
%with no particular regularity.
On the basis of this analogy, our intent
is to acquire an understanding of the wind dynamics through nonlinear
model equations.

To begin, let us consider the balance of forces (per mass) on the plume:
\begin{equation}
\label{force-balance}
\frac{dU}{dt}
=
f_{b}(\varphi,t)\,\sin\varphi
-
f_{d}.
\end{equation}
Substituting the previous relations into (\ref{force-balance}),
and integrating the resultant expression with respect to
$\varphi$ from $0$ to $2\pi$, one readily finds:
\begin{equation}
\label{U}
\frac{dU}{dt}
=
\frac{1}{2}\, \alpha_{p}\,g\,A_{1}
-
\frac{1}{2}
C(\Re)\,\frac{U^{2}}{L}.
\end{equation}
Remarkably, the temporal behavior of $U$ is coupled to
the amplitude of the first temperature mode $A_{1}$ only.

The temporal change of the plume temperature
is given by advection and by diffusion. For the latter,
we assume a relaxation ansatz for the temperature
deviation $T - T_{s}$ from the surrounding, with
the diffusive time scale $\tau_{\kappa} = \lambda_\theta L /\kappa$, i.e.,
\begin{equation}
\label{pde-T}
\frac{\partial T}{\partial t}
\;+\;
\frac{U}{L/2}\;\frac{\partial T}{\partial \varphi}
\;=\;
\,- \;\frac{T - T_{s}}{\tau_{\kappa}}.
\end{equation}
The physics behind the definition of $\tau_{\kappa}$ is that
the thermal loss is proportional to the plume surface, and
inversely proportional to the thermal diffusivity.

Substituting (\ref{T_surrounding}) and (\ref{T})
into (\ref{pde-T}), and equating the coefficients of each harmonic
separately, one obtains:
\begin{eqnarray}
\label{A}
\frac{dA_{1}}{dt} &=&
-\frac{\kappa}{\lambda_{\theta}L}\,A_{1}
+\frac{2}{L}UB_{1},\\
\label{B}
\frac{dB_{1}}{dt}
&=&
-\frac{\kappa}{\lambda_{\theta}L}\,B_{1}
+\frac{\kappa\Delta}{2\lambda_{\theta}L}\,
- \frac{2}{L}UA_{1}.
\end{eqnarray}
We write the three coupled ODEs (\ref{U}), (\ref{A}), and (\ref{B})
in nondimensional form. The dimensionless variables are
$X = 2^{-1}\, \Nu^{-2} \, \kappa^{-1}\,L\,U$,
$Y = 2\,r\,\Delta^{-1}\,A_{1}$,
$Z = (\Rac^{-1} - 2B_{1}\Delta^{-1} )r$,
$\tau = 2 \, \Nu \,\kappa\,L^{-2}\,t$, and
the dimensionless control parameters read:
\begin{equation}
\label{coefficients}
\sigma
=
\frac{9}{4}
\,
\frac{\Pr}{\Nu}, \qquad \mbox{and}\qquad
r
=
\frac{1}{18\Nu}\, \frac{\Ra}{\Rac},
\end{equation}
where $\Rac = 1708$, $\Pr = \nu/\kappa$ is the Prandtl number,
and $\Ra = \alpha_{p}gL^{3}\Delta/(\nu\kappa)$ the
Rayleigh number. The Nusselt number Nu comes from
the relation $\lambda_{\theta}/L = 1/(2\,\Nu)$.
Then, the system of equations (\ref{U}), (\ref{A})
and (\ref{B}) becomes:
\begin{eqnarray}
\label{turb-X}
\frac{dX}{d\tau} &=& \sigma\,Y
-
\sigma\,X
\left[
1
+
\sqrt{1 + \frac{27}{2b^{3}\sigma^{2}}\,X^{2}}
\right],
\\
\label{turb-Y}
\frac{dY}{d\tau} &=& r\,X - Y - X\,Z,\\
\label{turb-Z}
\frac{dZ}{d\tau} &=& -Z + X\,Y.
\end{eqnarray}
This system resembles the Lorenz equations
\cite{Lorenz/JAS/1963,Martin/PRA/1975}, which have also
been used to describe \textit{laminar} flow confined in a
toroidal loop \cite{Gorman/PRL/1984,Ehrhard/JFM/1990}.
Here equations (\ref{turb-X})-(\ref{turb-Z}) have been derived
to model plume reversals in the \textit{turbulent}
regime.
They will be referred to as the \textit{modified}
Lorenz equations. There are two essential differences
as compared to the standard Lorenz system \cite{footnote/1}:
(i)
The parameters $\sigma$ and $r$ are related to the
Nusselt number, which is known to follow a nonuniversal
(Pr-dependent) scaling with Re \cite{Grossmann/Lohse/GL}.
This is a key difference, since in the Lorenz equations
$\sigma = \Pr$ and $r = \Ra/\Rac$.
(ii)
The ordinary differential equation for $X$ has a \textit{new} nonlinear term,
due to the turbulent drag on the plume.

\textit{Phase diagram}:
To investigate the dynamical properties of the system
(\ref{turb-X})-(\ref{turb-Z}), we have scanned the parameter
space $\Ra \times \Pr$ in the range $10^{7} \leq \Ra \leq 10^{12}$,
$10^{-2} \leq \Pr \leq 10^{3}$. Technically, our numerical scheme
was based on a fourth-order Runge-Kutta method \cite{nrc},
with adaptive stepsize control in time, and increments of $0.1$
for $\log_{10}(\Pr)$ and $\log_{10}(\Ra)$.
The Nu-input required for coefficients (\ref{coefficients})
is provided by Grossmann-Lohse theory \cite{Grossmann/Lohse/GL},
and as initial condition we adopted $(X=1, Y=1, Z=1)$. As for
check with other initial values see later.

\begin{figure}[tbp]
\begin{center}
%\leavevmode
$\begin{array}{c}
\includegraphics*[width=6cm]{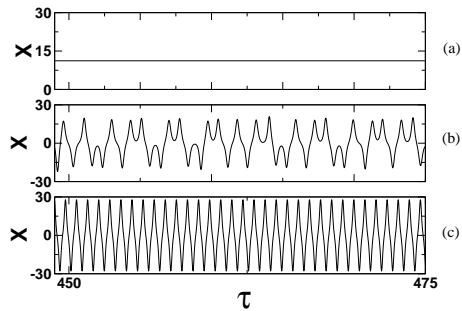}
\end{array}$
\end{center}
\vspace{-0.6cm}
\caption{\label{time_series}
Time series of the dimensionless plume velocity $X$,
for $\Pr = 0.1$ at
(a) $\Ra = 10^{10.8}$ (uniform circulation),
(b) $\Ra = 10^{10.9}$ (chaotic reversals), and
(c) $\Ra = 10^{11.0}$ (periodic reversals).}
\vspace{-0.5cm}
\end{figure}

\begin{figure}[tbp]
\begin{center}
$\begin{array}{c}
\\
\includegraphics*[width=7cm]{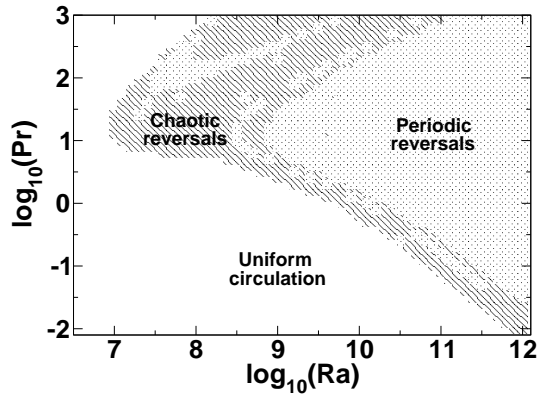}
\end{array}$
\end{center}
\vspace{-0.7cm}
\caption{\label{phase_diagram}
Phase diagram in the Ra $\times$ Pr plane: at sufficiently
large Ra, the state of uniform circulation (blank region) gives
place to chaotic (stripped region) or periodic (dotted region)
wind reversals. Note the small periodic windows in the chaotic
range.}
\vspace{-0.4cm}
\end{figure}

An insight into the structure of the phase diagram can
be acquired by considering some representative times series of $X(\tau)$.
In particular, for fixed Pr and increasing Ra, three
examples are shown in figure \ref{time_series}:
first, a state of uniform circulation [cf. plate (a)];
then emergence of chaotic reversals [plate (b)]; and, ultimately,
periodic reversals [plate (c)]. Figure \ref{phase_diagram} shows
the phase diagram in $\Ra \times \Pr$ space, displaying a sharp
onset between the steady and the reversal domain. We emphasize that
the transition curve between these domains remains unchanged for
a variety of initial conditions.

\textit{Onset of reversals}:
The onset of reversals can be understood in terms of the typical time
scales of the system: the thermal diffusion time
$\tau_{\kappa}$ and the turnover time $\tau_{U} = \pi L/\langle U \rangle$,
where $\langle \cdot \rangle$ denotes the time average.
Qualitatively, it is reasonable to expect wind reversals
when $\tau_{U} \ll \tau_{\kappa}$, because in such case
the circulation is so fast that the plume has no time to
lose its temperature contrast. Indeed, we find that the ratio
$\tau_{U}/\tau_{\kappa} = 2\pi \Nu\,\Pr^{-1}\,\langle \Re \rangle^{-1}$ is
a monotonically decreasing function of Ra for constant Pr, roughly
proportional to $\Ra^{-1/6}$.
%
%Its value $\tau_{U}/\tau_{\kappa}$ along
%the lower branch of the onset curve
%(between $10 > \Pr > 10^{-2}$ and $10^{7} \leq \Ra \leq 10^{12}$
%in fig. \ref{phase_diagram}) is essentially constant and equals 0.1.
%Along the upper branch (starting from $\Pr \geq 10$ and $\Ra \geq 10^{7}$)
%the ratio decreases as $\tau_{U}/\tau_{\kappa} \sim \Pr^{-0.40}$ or
%$\;\sim \Ra^{-0.55}$.
%
The overall form of the onset curve well resembles its
counterpart in the phase diagram of the Lorenz
model cf. Dullin\ \textit{et al.}
\cite{Dullin/Schmidt/Richter/Grossmann/2005}.

\begin{figure}[htbp]
\begin{center}
\includegraphics*[width=8.5cm]{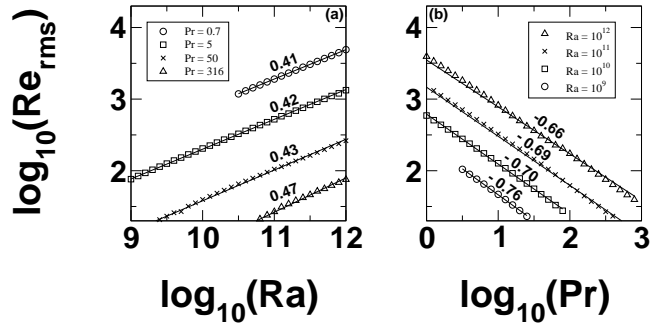}
\vspace{-0.6cm}
\end{center}
\caption{\label{Re_rms_Ra_Pr}
Reynolds number variance $\Re_{rms}$ based on the root mean square velocity
as function of (a) Ra and (b)  Pr.}
\vspace{-0.4cm}
\end{figure}

\textit{Reynolds number}:
We now come to the dependence of the variance of the Reynolds number
$\Re_{rms}  = L\,u_{rms}\,\nu^{-1}$ based on the root mean square velocity
$u_{rms} = \sqrt{\langle(U - \langle U \rangle)^{2}\rangle}$.
Figure \ref{Re_rms_Ra_Pr} shows $\Re_{rms}(\Ra,\Pr)$:
In plate (a), the Ra-scaling exponent increases from 0.41 to 0.47
for increasing Pr from 0.7 to 316; in plate (b), the Pr-scaling
exponent decreases from $-0.66$ to $-0.76$ for falling Ra from
$10^{12}$ to $10^{9}$. Experimentally, a similar Pr-dependence
has been reported \cite{Xia/PRE/2002} for the Reynolds numbers
based on the maximum wind velocity, on the oscillation frequency
of the large-scale circulation, and on the rms velocity.

\textit{Mean reversal frequency}:
The abrupt change of $X(\tau)$ with $\tau$ [cf. figure \ref{time_series}(b)]
suggests that the wind switching can be approximately considered as an
almost instantaneous event represented by the moment at which it occurs.
Here, we follow Sreenivasan \textit{et al.} \cite{Sreenivasan/wind/PRE/2002}
and define $t'_{n}$ as the interval between an arbitrary origin in time
and the \textit{n}th wind reversal. Similarly as in
\cite{Sreenivasan/wind/PRE/2002}, we also find a linear
relation $t'_{n} \sim n$, which suggests a mean interval $\bar{t'}$
between reversals. In this way, we define $\omega = 1/\bar{t'}$ as
the mean reversal frequency, and its dimensionless counterpart as
$\tilde{\omega} = \omega \, L^{2}/\nu$.

\begin{figure}[htbp]
\begin{center}
\includegraphics*[width=8.5cm]{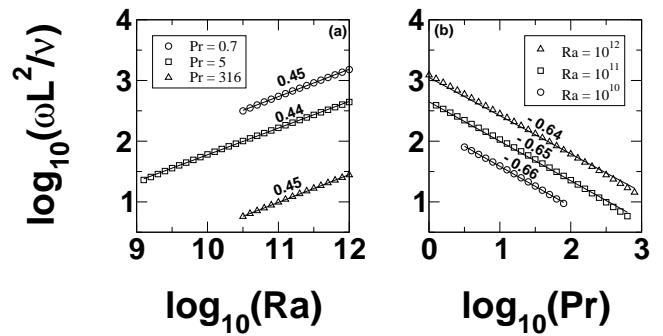}
\vspace{-0.4cm}
\end{center}
\caption{\label{frequency_Ra_Pr}
Mean reversal frequency as function of (a) Ra at fixed Pr and
(b) Pr for given Ra.}
\vspace{-0.2cm}
\end{figure}

Figure \ref{frequency_Ra_Pr} shows that
$\tilde{\omega} \sim \Pr^{-(0.65 \pm 0.01)}\,\Ra^{0.44 \pm 0.01}$.
To our knowledge, the only experimental measurement of the
reversal frequency has been carried out in cryogenic helium gas
\cite{Sreenivsan/Niemela/frequency/private}, namely
$\tilde \omega \sim  \Ra^{0.71}$ for $\Pr = 0.75$ and
$2.1 \times 10^{8} \leq \Ra \leq 1 \times 10^{13}$.
The disagreement between our result and the particular
measurement suggests that a model based on only 3-modes
for the plumes is quantitavely inadequate.
In this qualitative sense, our simple \textit{deterministic}
system well mimics the dynamics of reversals, and is a
complementary approach to the stochastic
model of noise-induced switchings between
two metastable states \cite{Sreenivasan/wind/PRE/2002}.
Here, the Lorenz attractor itself captures
the bistable transitions, but a more quantitative
description of the reversal phenomenon (also
including the rotation scenario) would involve a subtle
combination of deterministic chaos and noise. This could be done in
the spirit of ref. \cite{Silchenko/PRE/2002} (sec. III C),
and is left for future work.

\begin{acknowledgments}
We thank G.\ Ahlers, K.\ R.\ Sreenivasan, and J.\ Niemela
for fruitful exchange. This work is part of the research programme of
Stichting FOM, which is financially supported by NWO.

\end{acknowledgments}

%\vspace{-0.5cm}

%\vfill


\begin{thebibliography}{99}
\bibitem{footnote/1}
What however our modified Lorenz system has in common with the original
one is that it is not suited to make statements on the heat flux in the
turbulent regime.
\end{thebibliography}
\end{document}